# Complete phase retrieval of photoelectron wavepackets

L Pedrelli[1*], P D Keathley[2*], L Cattaneo[1], F X Kärtner[2,3], and U Keller[1]

[1] Physics Department, Institute of Quantum Electronics, ETH Zürich, 8093 Zürich, Switzerland
[2] Research Laboratory of Electronics, Massachusetts Institute of Technology, Cambridge, MA 02139, USA
[3] Center of Free-Electron Laser Science, DESY and Department of Physics, University of Hamburg, D-22607 Hamburg, Germany

E-mail: lucaped@phys.ethz.ch

* L. Pedrelli and P. D. Keathley contributed equally to this work.

**Abstract**

Coherent, broadband pulses of extreme ultraviolet (XUV) light provide a new and exciting tool for exploring attosecond electron dynamics. Using photoelectron streaking, interferometric spectrograms can be generated that contain a wealth of information about the phase properties of the photoionization process. If properly retrieved, this phase information reveals attosecond dynamics during photoelectron emission such as multielectron dynamics and resonance processes. However, until now, the full retrieval of the continuous electron wavepacket phase from isolated attosecond pulses has remained challenging. Here, after elucidating key approximations and limitations that hinder one from extracting the coherent electron wavepacket dynamics using available retrieval algorithms, we present a new method called Absolute Complex Dipole transmission matrix element reConstruction (ACDC). We apply the ACDC method to experimental spectrograms to resolve the phase and group delay difference between photoelectrons emitted from Ne and Ar. Our results reveal subtle dynamics in this group delay difference of photoelectrons emitted form Ar. These group delay dynamics were not resolvable with prior methods that were only able to extract phase information at discrete energy levels, emphasizing the importance of a complete and continuous phase retrieval technique such as ACDC. Here we also make this new ACDC retrieval algorithm available with appropriate citation in return.

## 1. Introduction

When high-energy photons interact with a gas target, photoelectrons are generated. While the rate of this photoelectron generation can be described by the gas species' cross-section, the attosecond dynamics of the emission process are more difficult to resolve [1]–[3]. Revealing these dynamics is a key challenge to understanding multi-electron interactions and resonant processes occurring in any system more complicated than the hydrogen atom [4]–[7]. However, the accurate characterization of photoionization dynamics requires the phase information acquired by the electron wavepacket during the transition from the ground state to the excited state in the ionization process.

There has been a large amount of interest in characterizing this phase information in the past years via reconstruction of attosecond beating by interference of two-photon transitions (RABBITT) [8], [9] or via a fully optical interferometric technique [10] which requires the use of discrete harmonics of an attosecond pulse train. Such methods lead to a coarse energy sampling and cannot resolve fast oscillations in the group delay as a function of energy. Recently, there has been interest in the analysis of continuous spectrograms from isolated attosecond pulses to increase the energy resolution of the relative phase of the emitted electron wavepackets [2], [3], [11], [12]. However, these methods are still encumbered by many approximations that limit their overall effectiveness.

In this work, we develop a new method for the complete characterization of photoelectron wavepackets using photoelectron streaking spectrograms from isolated attosecond pulses which we call Absolute Complex Dipole transmission matrix element reConstruction (ACDC). The ACDC method is not limited by any approximation other than the strong-field approximation (SFA) used to model the photoionization and streaking process, enabling unprecedented accuracy in the retrieval of the photoelectron wavepacket's phase across a continuous spectrum of electron energies. Applying ACDC to simultaneously measured photoelectron streaking spectrograms from Ar and Ne, we reveal subtle dynamics in the group delay of the photoionization process from Ar that would have been impossible to resolve using prior techniques that sampled the group delay at discrete energy points.

This paper is structured as follows. In section 2 we start with an overview of phase-retrieval techniques and their application in phase and group delay analysis of photoelectron wavepackets. In section 3 we introduce and briefly review the so-called wavepacket approximation [11], [13], that prevents conventional attosecond pulse characterization algorithms from accurately retrieving the complete electron wavepacket from streaking spectrograms. In section 4, we introduce the ACDC method, demonstrating its capability of overcoming the wavepacket approximation to fully characterize electron wavepackets and attosecond time delays in photoemission to within the limits of the SFA. Finally, in section 5, we use the ACDC method to reconstruct the dipole phase of electron wavepackets excited from Ar using experimental streaking measurements.

Our results agree very well with the average group delay measurements using discrete harmonics [10], showing almost a $\pi$ shift in phase at photon energies ranging from 26 to 30 eV, corresponding to the energy position of the well-known $3s3p^6np$ autoionizing series of Ar [14]. Furthermore, the ACDC reconstruction reveals new group-delay features between 30-35 eV that could not be resolved by other techniques. Our results emphasize the importance of the complete and continuous phase retrieval method we use here. We believe that, this work opens the door to more complex studies such as photoelectron excitation dynamics in molecular systems.

## 2. XUV phase retrieval techniques for photoelectron wavepacket analysis

To experimentally investigate attosecond photoemission dynamics, two methods are commonly used: (1) RABBITT [8], [9]; and (2) the photoelectron streaking from single attosecond pulses (SAPs) [15]. Both techniques employ a pump-probe scheme, where an extreme-ultraviolet (XUV) attosecond pump pulse initiates electron dynamics and an infrared (IR) probe pulse interrogates the temporal evolution of the released electrons as the delay τ between pump and probe is varied. While the RABBITT technique uses an attosecond pulse train (APT) in combination with a weak (<$10^{11}$ W/cm$^2$) and typically long IR (~30 fs) pulse, the photoelectron streaking method uses a SAP and a few-cycle IR (<6 fs) as a pump and probe, respectively.

In RABBITT, the spectrum of the applied APT only comprises odd multiples 2q+1 (q is an integer number) of the fundamental IR frequency $\omega_{IR}$ giving rise to a discrete photoelectron spectrum. Due to the presence of the weak IR probe field the ionized photoelectrons may absorb or emit one additional IR photon leading to so-called sidebands in between two consecutive harmonic energies. Due to the discrete nature of the resulting photoelectron spectrum, the retrieved phase is coarsely sampled by twice the fundamental photon energy ($2\hbar\omega_{IR}$ ~ 3eV, where $\hbar$ is the reduced Plank constant). A different and fully optical approach for characterizing electron wavefunctions and to extract photoionization dynamics information is described in [10] where the authors used an XUV-XUV time-resolved interferometry technique that has been recently proposed to extract the dipole phase difference between two atomic species from which the XUV attosecond pulse trains where generated [10]. However, this technique only allows sampling of the accumulated phase of the escaping electrons at energy points that are separated by $2\hbar\omega_{IR}$, i.e. at energy positions corresponding to the XUV harmonic peaks.

On the other hand, in the streaking measurements the photoelectrons are released upon the absorption of a XUV photon from a SAP in presence of a few-cycle IR pulse, which streaks the liberated electron wavepackets in the continuum to different final momenta. Since different original kinetic energy states are mapped to equivalent final kinetic energy states, there is interference of the electron pathways providing access to the original phase information of the photoemittted electron wavepacket, which can be retrieved with numerical analysis. Since the ionizing SAP contains all energies within its spectral bandwidth, the spectral phase of the electron wavepacket is sampled in a continuous manner, with no fundamental limitation on the energy sampling resolution.

The collected spectral phase of the photoionization and streaking process $\varphi_{stk}$ can be written as a sum of three main contributing terms [16], i.e.

$$\varphi_{stk} = \varphi_{XUV} + \varphi_W + \varphi_{CLC} \qquad (1)$$

where $\varphi_{XUV}$ is the attochirp phase of the SAP, $\varphi_W$ is the Wigner phase and $\varphi_{CLC}$ is the Coulomb-laser coupling phase which takes into account the extra phase that the photoemitted electron wavepacket acquires due to the interaction with both the long-range Coulomb potential and the IR streaking field, very similar to the continuum-continuum phase in the RABBITT technique [17].

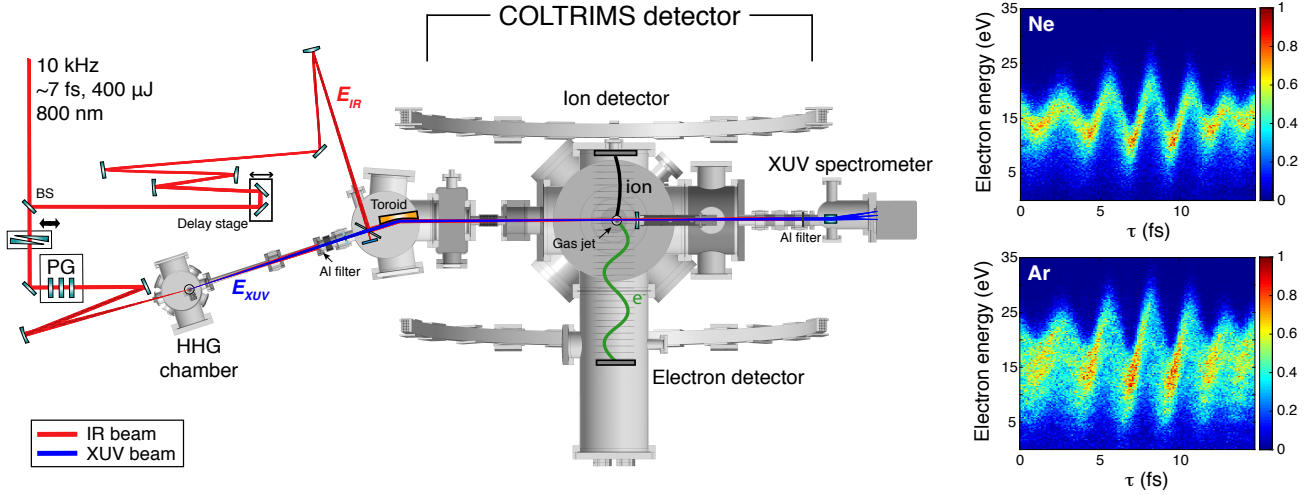

**Figure 1.** Scheme of the experimental setup including the COLTRIMS detector used to simultaneously measure the streaking spectrograms of Ne (top) and Ar (bottom) considered in this work. The red line represents the path of the IR beam which is spited in two parts by the beam splitter (BS). 80% is used to generate, using the polarization gating (PG) scheme, the XUV pump beam ($E_{XUV}$) while the remaining 20% is used as the probe streaking field ($E_{IR}$). The scheme of the experimental setup has been adapted from [19].

The derivative of the retrieved spectral phase $\varphi_{stk}(E)$ corresponds to the group delay (GD) of the excited electron wavepacket, *i.e.*

$$GD = \frac{\partial \varphi_{stk}(E)}{\partial E}. \quad (2)$$

The attochirp contribution to $\varphi_{stk}$ can be eliminated by computing the GD difference between two different species measured simultaneously and subject to the same ionizing XUV spectrum. This requires the use of a coincidence detection apparatus such as the Cold Target Recoil Ion Momentum Spectrometer (COLTRIMS) [18]. Experimental streaking spectrograms measured simultaneously in Ar and Ne using the COLTRIMS apparatus [19] and which will be object of our analysis in section 5 of this work are shown in Figure 1.

The measurement-induced contribution $\varphi_{CLC}$ can be extracted from numerical calculation [20] such that by comparison of two simultaneously measured streaking spectrograms the extracted phase has only the Wigner contribution. However, as alluded to earlier, access to $\varphi_W(E)$ from photoelectron streaking measurements requires a numerical retrieval algorithm.

Frequency resolved optical gating for complete retrieval of attosecond bursts (FROG-CRAB) [21], [22] is a well-established and widely used method for characterizing the time-domain profile of XUV pulses (both amplitude and phase) from streaking spectrograms. A critical analysis in retrieving the phase of the photoemitted electron wavepacket using FROG-CRAB methods has been carried out by Wei et al. [1] where they point out limitations of the method, namely the central momentum approximation and what they call the wavepacket approximation (WPA) [11], [13]. In their work, they were unable to overcome these limitations to fully recover the phase of the photoelectron wavepacket. While others have shown the ability to circumvent the central-momentum approximation [23], to our knowledge the wavepacket approximation remains the limiting factor in accurate photoelectron retrieval. In the following sections, we define the wavepacket approximation and demonstrate why it is so problematic. We then present the ACDC method which overcomes the wavepacket approximation and use it with the experimental data shown in Figure 1 to reveal newly observed group-delay dynamics in photoemission from Ar.

## 3. Wavepacket approximation (WPA)

The complex amplitude of a photoelectron going from the ground state to the final state with momentum $k$ can be expressed within the SFA

$$\tilde{a}_{SFA}(k,\tau) = -i \int_{-\infty}^{+\infty} dt \tilde{d}(k+A(t)) \\ \tilde{E}_{XUV}(t-\tau) e^{i\phi(k,t)} e^{i\left(I_P + \frac{k^2}{2}\right)t}, \quad (3)$$

where $\tilde{d}(k)$ is the dipole transition matrix element (DTME), $\tau$ the time delay between the XUV pump pulse and the IR probe pulse, $\tilde{E}_{XUV}(t)$ is the XUV field in time domain and $A(t)$ the vector potential of the IR electric field from which it can be calculated, $E_{IR}(t) = -\frac{\partial}{\partial t}A(t)$.

The phase modulation term, $\phi(k,t)$, resulting from the accumulated phase of the electron during the streaking process reads

$$\phi(k,t) = -\int_t^{+\infty} dt' \left[kA(t') + \frac{1}{2}A^2(t')\right]. \quad (4)$$

The measured photoelectron spectrum is given by the probability of measuring an electron with momentum $k$ and delay $\tau$, *i.e.* $S_{SFA}(k,\tau) = |\tilde{a}_{SFA}(k,\tau)|^2$.

In absence of the IR electric field, the XUV pulse, ionizing the target gas, generate a photoelectron wavepacket that can be expressed in the energy domain as

$$\tilde{\chi}(E) = \tilde{E}_{\text{XUV}}\left(I_P + \frac{k^2}{2}\right)\tilde{d}(E). \quad (5)$$

Introducing the expression of the wavepacket (5) into equation (3) we can rewrite the SFA complex amplitude within the WPA as follows

$$\tilde{a}_{\text{WPA}}(k,\tau) = -i\int_{-\infty}^{+\infty} dt\,\tilde{\chi}(t-\tau)e^{i\phi(k,t)}e^{i\left(I_P+\frac{k^2}{2}\right)t}. \quad (6)$$

As explained in [11], the WPA is not an exact theory but rather an approximation to the SFA model given by equation (3), which is considered the model better describing the physics of the streaking process. The difference between the spectrogram $S_{\text{SFA}}(k,\tau)$ and $S_{\text{WPA}}(k,\tau) = |\tilde{a}_{\text{WPA}}(k,\tau)|^2$ can be seen by comparing $S_{\text{SFA}}$ and $S_{\text{WPA}}$ as we show in Figure 2.

Most photoelectron retrieval methods based on FROG-CRAB, when used to characterize the electron wavepacket $\tilde{\chi}(E)$, rely on the WPA (i.e. they use $\tilde{a}_{\text{WPA}}$ to model the streaking process rather than $\tilde{a}_{\text{SFA}}$), which leads to significant errors in the retrieved phase response and group delay as the physics of the experiment is better described by $\tilde{a}_{\text{SFA}}$ (see appendix A for further analysis). In the following sections, we introduce the ACDC method which overcomes the WPA, enabling a complete characterization of the photoemitted electron wavepacket.

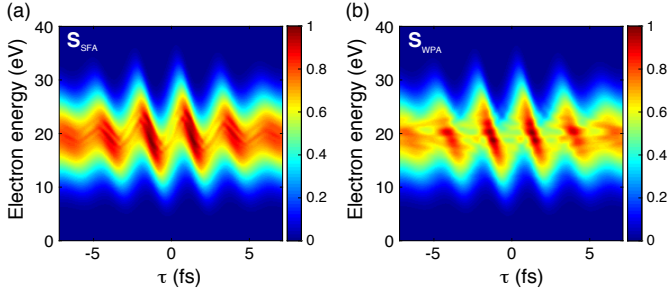

**Figure 2.** Simulated streaking spectrograms using (a) the SFA model $S_{\text{SFA}}(k,\tau) = |\tilde{a}_{\text{SFA}}(k,\tau)|^2$ and (b) using the WPA, i.e. $S_{\text{WPA}}(k,\tau) = |\tilde{a}_{\text{WPA}}(k,\tau)|^2$. We observe a percentage difference between the two spectrograms shown here of up to 35% mainly concentrated around the central energy of the spectrogram.

## 4. Absolute Complex DTME reConstruction (ACDC) algorithm

In this section we describe the method of the ACDC algorithm to retrieve the DTME ($\tilde{d}$), when both the XUV and IR pulses are known quantities.

Starting from the SFA expression of the complex amplitude of the photoemitted electron wavepacket given by equation (3), we now approximate $\tilde{d}$ using the first order Taylor expansion at such that

$$\tilde{d}(k + A(t+\tau)) \approx \tilde{d}(k) + \tilde{d}'(k)A(t+\tau), \quad (7)$$

where $\tilde{d}'(k) = \partial_k \tilde{d}$. In this way we can rewrite the complex amplitude describing the transition from the ground state to the final momentum $k$ as:

$$\tilde{a}_{\text{SFA}}(k,\tau) \approx \tilde{d}(k)\tilde{\Gamma}(k,\tau) + \tilde{d}'(k)\tilde{\beta}(k,\tau) \quad (8)$$

where we introduced

$$\tilde{\Gamma}(k,\tau) = -i\int_{-\infty}^{+\infty} dt\,\tilde{E}_X(t)e^{i\phi(k,t+\tau)}e^{i\left(I_P t+\frac{k^2 t}{2}\right)} \quad (9)$$

and

$$\tilde{\beta}(k,\tau) = -i\int_{-\infty}^{+\infty} dt\,A(t+\tau)\tilde{E}_X(t)e^{i\phi(k,t+\tau)}e^{i\left(I_P t+\frac{k^2 t}{2}\right)}. \quad (10)$$

When we numerically sample the momentum and time delay components, we approximate the derivative of $\tilde{d}$ as:

$$\tilde{d}'[m] = \frac{\tilde{d}[m+1] - \tilde{d}[m-1]}{\Delta k[m]} \quad (11)$$

where $m$ is the sample number in energy and $\Delta k[m] = k[m+1] - k[m-1]$. Using this we can write

$$\tilde{a}_{\text{SFA}}[l,m] = \tilde{d}[m]\tilde{\Gamma}[l,m] + \frac{\tilde{d}[m+1]\tilde{\beta}[l,m]}{\Delta k[m]} - \frac{\tilde{d}[m-1]\tilde{\beta}[l,m]}{\Delta k[m]} \quad (12)$$

where $l$ is the sample number of time delays.

Using the same minimization strategy as Volkov transform generalized projections algorithm (VTGPA) [23] we define the matrix $\tilde{a}'[l,m]$ as

$$\tilde{a}'[l,m] = \sqrt{P[l,m]}e^{i\,\arg(\tilde{a}_{\text{SFA}}[l,m])} \quad (13)$$

that is the measured spectrogram amplitude $P[l,m]$ projected onto the computed spectrogram $\tilde{a}_{\text{SFA}}[l,m]$. From this we can define the figure of merit $M$

$$M = \sum_l \sum_m (\tilde{a}[l,m] - \tilde{a}'[l,m]) \times \text{c.c.} \quad (14)$$

and the value of $\tilde{d}[m] = d[m]e^{i\phi[m]}$ that minimizes $M$ by solving the system of equations

$$\frac{\partial M}{\partial d[m]} = 0; \qquad \frac{\partial M}{\partial \phi[m]} = 0. \quad (15)$$

In appendix B we report the full derivation of the expression of $\tilde{d}$.

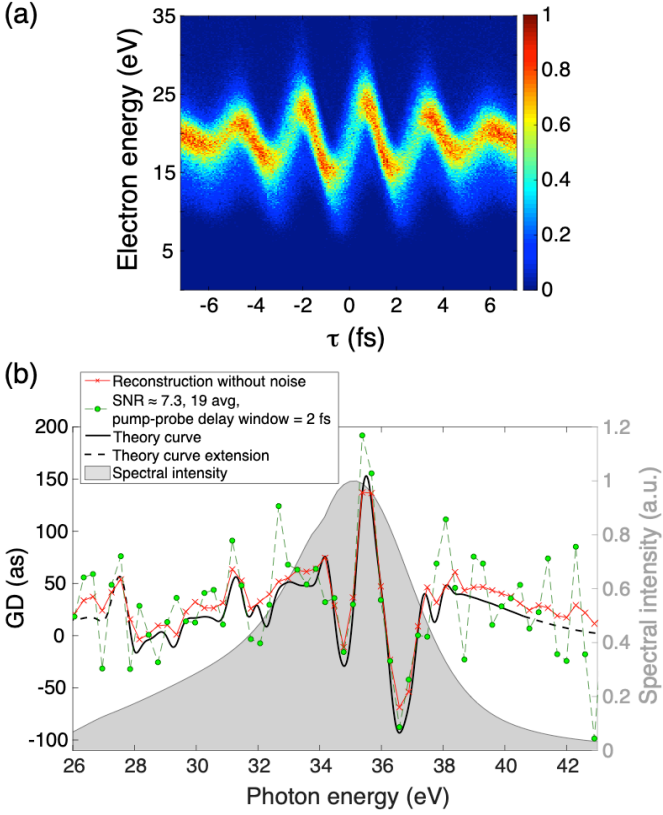

**Figure 3.** (a) Simulated streaking spectrogram with a Poisson noise level comparable to the experimental conditions. (b) Retrieved GD with ACDC without noise (red line) and in presence of noise (green line). The green line is the weighted average of 19 reconstructions on different pump probe delay windows of 2 fs. The theory curve (black line) used as input to generate the simulated spectrogram shown in (a) is from reference [3]. In order to be able to simulate the spectrograms avoiding unphysical abrupt jumps in the simulated spectrogram amplitude, we extended the theoretical curve below 27.8 eV and above 40.5 eV (black dashed lines). The grey shaded area shows the XUV spectral intensity reproducing the experimental case.

For low streaking intensities ($\leq 1\times10^{10}$ W/cm$^2$), this approach accurately reconstructs $\tilde{d}$ (see simulation results in Appendix B). However, when the streaking intensity, and thus the streaking vector potential, is large ($> 1\times10^{10}$ W/cm$^2$), the Taylor expansion used in equation (7) becomes inadequate. Expanding to higher orders was found to be overly cumbersome, and in the end the most effective approach in this case was to: (1) use the analytical method described above using the first order Taylor expansion as a first step to estimate the amplitude and phase of $\tilde{d}$; and (2) apply a numerical stochastic gradient descent algorithm to refine the estimated DTME from (1), bringing it much closer to the actual solution.

For step (2), we used a stochastic gradient descent algorithm along with Adam (Adaptive Momentum Estimation) optimizer [24]. The use of step (1) is justified by the fact that, the convergence to local minima of the cost function, when using the step (2), results faster.

For simplicity, we leave the technical description, together with the effect of the approximation of the first order Taylor expansion, in appendix B. Additionally, in appendix C we show the robustness of the ACDC method on simulated spectrograms using different streaking parameters, *i.e.* IR intensity, XUV bandwidth and XUV chirp.

In Figure 3, we show the result of the ACDC method when reconstructing a simulated spectrogram from Ar. Artificial Poisson noise was added to the simulated spectrogram to mimic the noise observed in the experimental data (Fig. 1). Since we want to reconstruct the DTME from experimental spectrograms, we also considered parameters of both the XUV and the IR field similar to the experimental conditions that we will consider in the next section. In particular, we used an IR intensity of $3\times10^{12}$ W/cm$^2$ while the XUV spectrum (grey shaded area in Figure 3) resembles the actual spectrum used during experiments. The red line in Figure 3 represents the reconstructed GD using the ACDC method in absence of noise while the green curve shows the result in presence of noise level comparable to the experiment (see appendix D for details of noise analysis). Despite the fact that fluctuations due to noise are still present, we observe a good agreement with the theoretical curve used as the input for the GD$^{Ar}$ (black line) [3] over the entire energy window considered for the reconstruction, including the rapid group delay fluctuations from 34-38 eV. This result validates the algorithm's robustness against noise and its applicability to experimental measurements which is considered in the next section.

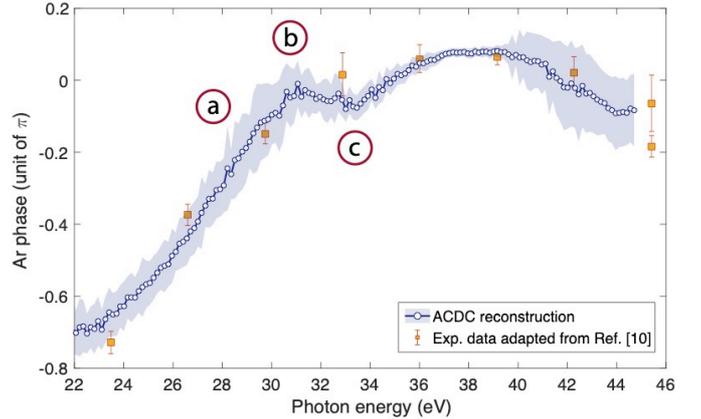

**Figure 4.** The blue curve displays the average of 36 retrieved Ar dipole phases from the reconstruction of experimental measurement using the ACDC algorithm. The yellow symbols are adapted experimental result from reference [10] and shifted to within a constant offset. The labels (a), (b) and (c) refer to the discussion in the main text.

## 5. Experimental reconstruction of the Ar dipole phase

In this section we retrieved the dipole phase of Ar from the experimental spectrogram shown in Figure 1 using ACDC. In Figure 4 the blue line and blue shaded area show the reconstructed Ar dipole phase resulting from 36 averages using the ACDC algorithm and its standard deviation, respectively. For the interested reader, the details of the reconstruction procedure can be found in appendix E.

The dipole phase difference between Ne and Ar have also been measured experimentally by Azoury et al. using optical attosecond interferometry [10]. Adapting their result by flipping the sign and adding the Ne dipole phase from reference [25] (yellow squares in Figure 4), we observe an excellent agreement with the ACDC reconstruction over the full energy spectrum. However, since the ACDC algorithm resolves the phase in a continuous manner, a much finer phase structure is resolved as a function of photon energy.

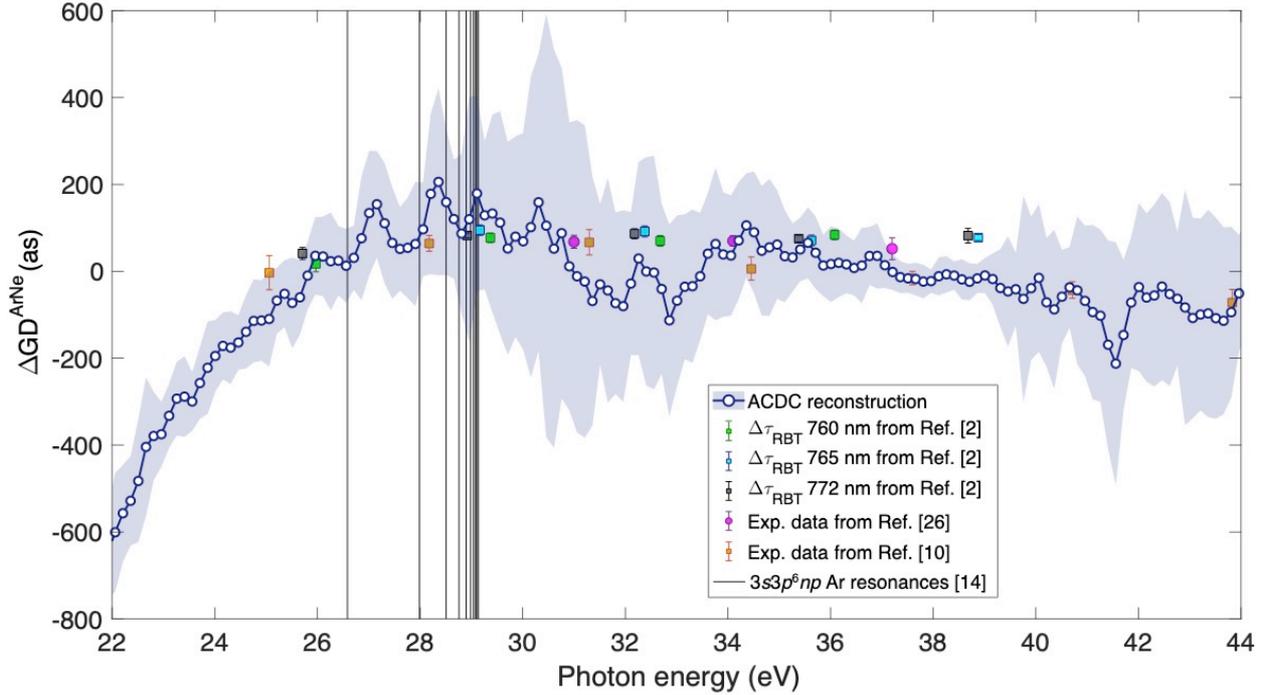

**Figure 5.** Retrieved $\Delta GD^{ArNe}$ from experimental streaking measurement using the ACDC algorithm averaging 36 reconstructions is shown by the blue curve together with its standard deviation (blue shaded area). RABBITT measurements from reference [2] (green, blue and grey squares) and from reference [26] (magenta circles) are displayed. Results from a fully optical interferometric technique from reference [10] are also plotted with yellow squares. The black vertical lines show the calculated $3s3p^6np$ autoionizing series of Ar from reference [14].

In the present measurement we sampled the Ar DTME every 0.15 eV, while in [10] it was every ~3.1 eV due to the spacing of the discrete XUV harmonics.

The ACDC result reveals a pronounced peak in the dipole phase of Ar between 30 to 32 eV (label (b) in Figure 4) and a subsequent dip in the energy region from 33 to 35 eV (label (c) in Figure 4) which cannot be described by a piecewise linear fit between the data points from [10].

In order to compare the obtained $GD^{Ar}$ with the differential group delay $\Delta GD^{ArNe}$ reported from other experiments, we subtracted the GD of Ne from $GD^{Ar}$ using the theoretical calculation from reference [25]. The result is shown in Figure 5 by the blue curve together with the blue shaded region, which represents the standard deviation. We note the persistent rapid variations of the $\Delta GD^{ArNe}$ that lie in the energy region where the Ar autoionizing states are located as displayed by the black vertical lines. While these oscillations are intriguing, they are near the noise limits of the retrievals from our current measurement data, and to fully confirm that those oscillations are signature of the $3s3p^6np$ autoionizing states will require further investigations that go beyond the scope of this work.

In Figure 5 we also plot several data points from previously reported RABBITT-like measurements of the average group delay $\Delta GD^{ArNe}$ (green, blue and grey squares [2], magenta circles [26], and yellow squares [10]). We observe that, between 28 to 30 eV the ACDC result retrieves, on average, a higher $\Delta GD^{ArNe}$ compared to RABBITT measurements while, between 30 to 34 eV, the ACDC result predicts a lower value. These correspond to the regions labelled as (a), (b) and (c) in Figure 4. This discrepancy between $\Delta GD^{ArNe}$ retrieved by ACDC and the RABBITT measurements arises due to the ACDC's ability to more finely sample the phase of the DTME. Indeed, the comparison between the $\Delta GD^{ArNe}$ curve reconstructed with the ACDC from the SAP streaking measurement and the other data points derived from discrete energy measurements using the RABBITT-like techniques will never perfectly overlap wherever the phase is not well-fit by a piecewise linear phase function between the RABBITT sampling points. For example, after downsampling the reconstructed phase of Ar using the ACDC algorithm from Figure 4 at the energy points separated by ~3.1 eV of the experimental work presented in reference [10] (yellow squares) and then computing the $\Delta GD^{ArNe}$, we obtain the green diamonds shown in Figure 6 which overlap almost perfectly with the datapoints from reference [10] (yellow squares).

## 6. Conclusion

In this work, we have addressed long-standing issues in the characterization of electron wavepackets using streaking spectrograms. We demonstrated that the WPA hinders the accurate retrieval of an electron wavepacket when using conventional attosecond pulse characterization algorithms. To overcome this approximation and to exploit the continuous energy resolution of the photoelectron streaking technique, we introduced a new method for electron wavepacket characterization: ACDC. This algorithm has the ability to extract the electron wavepacket phase, thus the photoemission GD, within the limitations only imposed by the SFA, using well-established attosecond streaking methods.

We used the new algorithm to reconstruct the absolute GD of Ar from experimental measurements. The result highlights the importance of reconstructing the dipole phase in a continuous manner in order to resolve variations in the GD as a function of energy which would not be resolved by RABBITT-like techniques. Interestingly, the ACDC reconstruction reveals features in the $\Delta GD^{ArNe}$ in the energy region corresponding to that of the $3s3p^6np$ autoionizing series of Ar however a deeper investigation of this energy region is required using additional measurements and theoretical modeling.

We believe that with the ACDC algorithm, we can now fully exploit the potential offered by photoelectron streaking using SAPs. Specifically, we can now access the complete and continuous electron wavepacket phase profile of an unknown target assuming that the DTME of one target gas species is known, allowing the direct measurement of electron dynamics in more complicated systems, such as molecules. Experimental effort in this direction is currently underway.

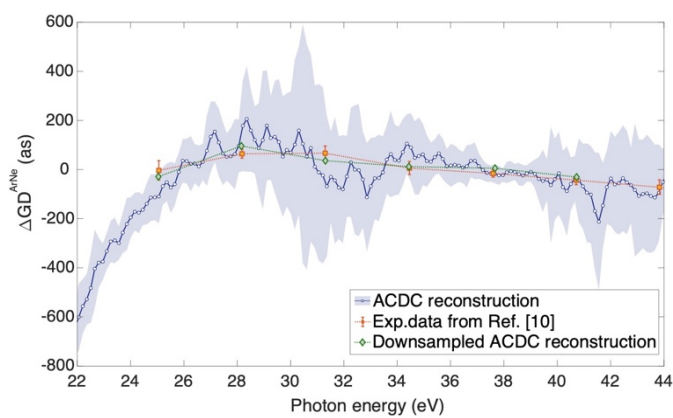

**Figure 6.** The green diamonds show the retrieved $\Delta GD^{ArNe}$ computed from the reconstructed dipole phase of Ar using the ACDC algorithm (blue curve in Figure 4) downsampled at the energy points separated by ~3.1 eV of the experimental result presented in reference [10] (yellow squares).

## Acknowledgements

This work was supported by NCCR Molecular Ultrafast Science and Technology (NCCR MUST), research instrument of the Swiss National Science Foundation (SNSF).

P. Keathley acknowledges support by the Air Force Office of Scientific Research (AFOSR) grant under contract NO. FA9550-19-1-0065.

SUPPLEMENTARY INFORMATION

## Appendix A: Effect of the WPA on characterization of electron wavepacket

We reconstructed the $S_{SFA}$ shown in Figure 2 (a) using the VTGPA which requires the application of the WPA due to the fact that the DTME is not known apriori. The retrieved dipole phase and the corresponding GD are shown by the blue curves in Figure S 1 (a) and (b), respectively. The deviation from the theory curve (black lines) used to simulate the $S_{SFA}$ shows the effect introduced by the WPA when trying to reconstruct the electron wavepacket using conventional algorithms. In general, the WPA leads to a smoothing of the retrieved phase profile as a function of energy. Similar errors are reported in [1].

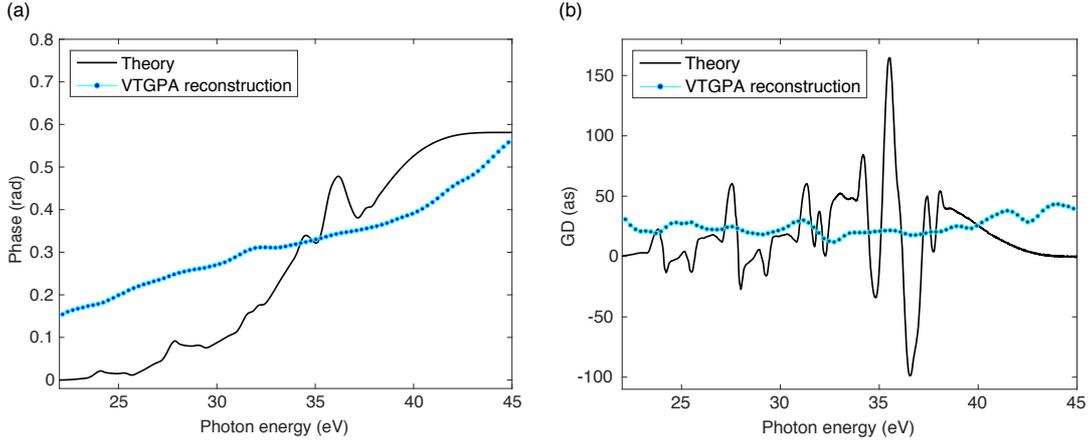

**Figure S 1.** Retrieved (a) phase and (b) GD using the VTGPA algorithm for the $S_{SFA}$ shown in Figure 2. The black curves represent the theory curves for the phase and GD used to simulate $S_{SFA}$.

## Appendix B: Full derivation of the ACDC algorithm

Continuing from section 4 we present here the full derivation of $\tilde{d}$. We can write the full expression of $M$ from equation (14) using the expression of $\tilde{a}[l,m]$ (equation (12)). This results in:

$$M = \sum_l \sum_m d[m]^2 \Gamma[l,m]^2 + \frac{d[m+1]^2 \beta[l,m]^2}{\Delta k[m]^2} + \frac{d[m-1]^2 \beta[l,m]^2}{\Delta k[m]^2} + a[l,m]^2 \\ + 2\Re\left( \tilde{d}[m]\tilde{\Gamma}[l,m] \frac{\tilde{d}[m+1]^* \tilde{\beta}[l,m]^*}{\Delta k[m]} - \tilde{d}[m]\tilde{\Gamma}[l,m] \frac{\tilde{d}[m-1]^* \tilde{\beta}[l,m]^*}{\Delta k[m]} \right. \\ \left. - \tilde{d}[m]\tilde{\Gamma}[l,m]\tilde{a}'[l,m]^* - \frac{\tilde{d}[m+1]\tilde{d}[m-1]^* \beta[l,m]^2}{\Delta k[m]^2} - \frac{\tilde{d}[m+1]\tilde{\beta}[l,m]\tilde{a}'[l,m]^*}{\Delta k[m]} \right. \\ \left. + \frac{\tilde{d}[m-1]\tilde{\beta}[l,m]\tilde{a}'[l,m]^*}{\Delta k[m]} \right) \tag{S 1}$$

We consider the derivative of this expression with respect to the term $m = c$. The only terms of $M$ that matter are $M_c$, $M_{c+1}$ and $M_{c-1}$. These terms are defined in the following equations

$$M_c = \sum_l d[c]^2 \Gamma[l,c]^2 + \frac{d[c+1]^2 \beta[l,c]^2}{\Delta k[c]^2} + \frac{d[c-1]^2 \beta[l,c]^2}{\Delta k[c]^2} + a[l,c]^2$$
$$+ 2\Re\left(\tilde{d}[c]\tilde{\Gamma}[l,c]\frac{\tilde{d}[c+1]^*\tilde{\beta}[l,c]^*}{\Delta k[c]} - \tilde{d}[c]\tilde{\Gamma}[l,c]\frac{\tilde{d}[c-1]^*\tilde{\beta}[l,c]^*}{\Delta k[c]}\right.$$
$$- \tilde{d}[c]\tilde{\Gamma}[l,c]\tilde{a}'[l,c]^* - \frac{\tilde{d}[c+1]\tilde{d}[c-1]^*\beta[l,c]^2}{\Delta k[c]^2} - \frac{\tilde{d}[c+1]\tilde{\beta}[l,c]\tilde{a}'[l,c]^*}{\Delta k[c]}$$
$$\left.+ \frac{\tilde{d}[c-1]\tilde{\beta}[l,c]\tilde{a}'[l,c]^*}{\Delta k[c]}\right) \quad \text{(S 2)}$$

$$M_{c+1} = \sum_l d[c+1]^2 \Gamma[l,c+1]^2 + \frac{d[c+2]^2 \beta[l,c+1]^2}{\Delta k[c+1]^2} + \frac{d[c]^2 \beta[l,c+1]^2}{\Delta k[c+1]^2} + a[l,c+1]^2$$
$$+ 2\Re\left(\tilde{d}[c+1]\tilde{\Gamma}[l,c+1]\frac{\tilde{d}[c+2]^*\tilde{\beta}[l,c+1]^*}{\Delta k[c+1]}\right.$$
$$- \tilde{d}[c+1]\tilde{\Gamma}[l,c+1]\frac{\tilde{d}[c]^*\tilde{\beta}[l,c+1]^*}{\Delta k[c+1]} - \tilde{d}[c+1]\tilde{\Gamma}[l,c+1]\tilde{a}'[l,c+1]^*$$
$$- \frac{\tilde{d}[c+2]\tilde{d}[c]^*\beta[l,c+1]^2}{\Delta k[c+1]^2} - \frac{\tilde{d}[c+2]\tilde{\beta}[l,c+1]\tilde{a}'[l,c+1]^*}{\Delta k[c+1]}$$
$$\left.+ \frac{\tilde{d}[c]\tilde{\beta}[l,c+1]\tilde{a}'[l,c+1]^*}{\Delta k[c+1]}\right) \quad \text{(S 3)}$$

$$M_{c-1} = \sum_l d[c-1]^2 \Gamma[l,c-1]^2 + \frac{d[c]^2 \beta[l,c-1]^2}{\Delta k[c-1]^2} + \frac{d[c-2]^2 \beta[l,c-1]^2}{\Delta k[c-1]^2} + a[l,c-1]^2$$
$$+ 2\Re\left(\tilde{d}[c-1]\tilde{\Gamma}[l,c-1]\frac{\tilde{d}[c]^*\tilde{\beta}[l,c-1]^*}{\Delta k[c-1]}\right.$$
$$- \tilde{d}[c-1]\tilde{\Gamma}[l,c-1]\frac{\tilde{d}[c-2]^*\tilde{\beta}[l,c-1]^*}{\Delta k[c-1]} - \tilde{d}[c-1]\tilde{\Gamma}[l,c-1]\tilde{a}'[l,c-1]^*$$
$$- \frac{\tilde{d}[c]\tilde{d}[c-2]^*\beta[l,c-1]^2}{\Delta k[c-1]^2} - \frac{\tilde{d}[c]\tilde{\beta}[l,c-1]\tilde{a}'[l,c-1]^*}{\Delta k[c-1]}$$
$$\left.+ \frac{\tilde{d}[c-2]\tilde{\beta}[l,c-1]\tilde{a}'[l,c-1]^*}{\Delta k[c-1]}\right) \quad \text{(S 4)}$$

From these expressions we need to solve the system of equations (15). If we consider the derivative respect to the magnitude $d[c]$ of the complex DTME we obtain:

$$\frac{\partial M_c}{\partial d[c]} = \sum_l 2d[c]\Gamma[l,c]^2$$
$$+ 2\Re\left(e^{i\phi[c]}\tilde{\Gamma}[l,c]\frac{\tilde{d}[c+1]^*\tilde{\beta}[l,c]^*}{\Delta k[c]} - e^{i\phi[c]}\tilde{\Gamma}[l,c]\frac{\tilde{d}[c-1]^*\tilde{\beta}[l,c]^*}{\Delta k[c]}\right.$$
$$\left.- e^{i\phi[c]}\tilde{\Gamma}[l,c]\tilde{a}'[l,c]^*\right) \quad \text{(S 5)}$$

$$\frac{\partial M_{c+1}}{\partial d[c]} = \sum_l \frac{2d[c]\beta[l,c+1]^2}{\Delta k[c+1]^2}$$
$$+ 2\Re\left(-e^{i\phi[c]}\tilde{d}[c+1]^*\tilde{\Gamma}[l,c+1]^*\frac{\tilde{\beta}[l,c+1]}{\Delta k[c+1]} - e^{i\phi[c]}\frac{\tilde{d}[c+2]\beta[l,c+1]^2}{\Delta k[c+1]^2}\right.$$
$$\left. + e^{i\phi[c]}\frac{\tilde{\beta}[l,c+1]\tilde{a}'[l,c+1]^*}{\Delta k[c+1]}\right)$$
(S 6)

$$\frac{\partial M_{c-1}}{\partial d[c]} = \sum_l \frac{2d[c]\beta[l,c-1]^2}{\Delta k[c-1]^2}$$
$$+ 2\Re\left(e^{i\phi[c]}\tilde{d}[c-1]^*\tilde{\Gamma}[l,c-1]^*\frac{\tilde{\beta}[l,c-1]}{\Delta k[c-1]} - e^{i\phi[c]}\frac{\tilde{d}[c-2]^*\beta[l,c-1]^2}{\Delta k[c-1]^2}\right.$$
$$\left. - e^{i\phi[c]}\frac{\tilde{\beta}[l,c-1]\tilde{a}'[l,c-1]^*}{\Delta k[c-1]}\right)$$
(S 7)

If we now introduce the following expressions:

$$\tilde{\delta}[l,c] = \tilde{\Gamma}[l,c]\frac{\tilde{d}[c+1]^*\tilde{\beta}[l,c]^*}{\Delta k[c]} - \tilde{\Gamma}[l,c]\frac{\tilde{d}[c-1]^*\tilde{\beta}[l,c]^*}{\Delta k[c]} - \tilde{\Gamma}[l,c]\tilde{a}'[l,c]^* \quad \text{(S 8)}$$

$$\tilde{\eta}[l,c] = -\tilde{d}[c+1]^*\tilde{\Gamma}[l,c+1]^*\frac{\tilde{\beta}[l,c+1]}{\Delta k[c+1]} - \frac{\tilde{d}[c+2]\beta[l,c+1]^2}{\Delta k[c+1]^2} + \frac{\tilde{\beta}[l,c+1]\tilde{a}'[l,c+1]^*}{\Delta k[c+1]} \quad \text{(S 9)}$$

$$\tilde{\gamma}[l,c] = \tilde{d}[c-1]^*\tilde{\Gamma}[l,c-1]^*\frac{\tilde{\beta}[l,c-1]}{\Delta k[c-1]} - \frac{\tilde{d}[c-2]^*\beta[l,c-1]^2}{\Delta k[c-1]^2} - \frac{\tilde{\beta}[l,c-1]\tilde{a}'[l,c-1]^*}{\Delta k[c-1]} \quad \text{(S 10)}$$

Using these expressions, we can write a simplified formula for $\frac{\partial M}{\partial d[c]}$ which reads

$$\frac{\partial M}{\partial d[c]} = \sum_l 2d[c]\Gamma[l,c]^2 + \frac{2d[c]\beta[l,c+1]^2}{\Delta k[c+1]^2} + \frac{2d[c]\beta[l,c-1]^2}{\Delta k[c-1]^2}$$
$$+ 2\Re\left(e^{i\phi[c]}(\tilde{\gamma}[l,c] + \tilde{\delta}[l,c] + \tilde{\eta}[l,c])\right)$$
(S 11)

With the same idea we can find the expression of $\frac{\partial M}{\partial \phi[c]}$ which results in

$$\frac{\partial M}{\partial \phi[c]} = \sum_l 2\Re\left(id[c]e^{i\phi[c]}(\tilde{\gamma}[l,c] + \tilde{\delta}[l,c] + \tilde{\eta}[l,c])\right) \quad \text{(S 12)}$$

that can be rewritten in the following way

$$\frac{\partial M}{\partial \phi[c]} = \sum_l -2d[c]\Im\left(e^{i\phi[c]}(\tilde{\gamma}[l,c] + \tilde{\delta}[l,c] + \tilde{\eta}[l,c])\right) \tag{S 13}$$

To solve the system of equations (15) we set both the equations (S 11) and (S 13) equal to zero. After some manipulation one get

$$0 = \sum_l 2d[c]\Gamma[l,c]^2 + \frac{2d[c]\beta[l,c+1]^2}{\Delta k[c+1]^2} + \frac{2d[c]\beta[l,c-1]^2}{\Delta k[c-1]^2} \\ + 2\Re\left(e^{i\phi[c]}(\tilde{\gamma}[l,c] + \tilde{\delta}[l,c] + \tilde{\eta}[l,c])\right) + 2i\Im\left(e^{i\phi[c]}(\tilde{\gamma}[l,c] + \tilde{\delta}[l,c] + \tilde{\eta}[l,c])\right) \tag{S 14}$$

The real and imaginary part are the same, so we can just break it out and rewrite (S 14) as

$$0 = \sum_l 2d[c]\Gamma[l,c]^2 + \frac{2d[c]\beta[l,c+1]^2}{\Delta k[c+1]^2} + \frac{2d[c]\beta[l,c-1]^2}{\Delta k[c-1]^2} + 2e^{i\phi[c]}(\tilde{\gamma}[l,c] + \tilde{\delta}[l,c] + \tilde{\eta}[l,c]) \tag{S 15}$$

We can finally solve for $\tilde{d}[n] = d[n]e^{i\phi[n]}$ which results in the final expression

$$\tilde{d}[c] = \frac{-\left(\sum_l \tilde{\gamma}[l,c]^* + \tilde{\delta}[l,c]^* + \tilde{\eta}[l,c]^*\right)}{\sum_l \Gamma[l,c]^2 + \frac{\beta[l,c+1]^2}{\Delta k[c+1]^2} + \frac{\beta[l,c-1]^2}{\Delta k[c-1]^2}} \tag{S 16}$$

The output result is then refined by using a stochastic gradient descent algorithm. Starting from the complex DTME $\tilde{d}$ solved by the ACDC algorithm, we use the gradient descent algorithm to further minimize the figure of merit. First, the gradient descent algorithm adds an arbitrarily small quantity $\Delta\phi$ to the $m^{\text{th}}$ energy point of the DTME phase

$$\phi_{\text{in}}[m] = \phi_{\text{in}}[m] + \Delta\phi. \tag{S 17}$$

The error associated to the spectrogram generated by the updated DTME function ($\varepsilon_{\text{new}}$) is compared with the starting error ($\varepsilon_{\text{start}}$). This provides an estimation of the gradient of the error function $\mathcal{E}$ that we want to minimize

$$\nabla\mathcal{E} = \frac{\varepsilon_{\text{new}} - \varepsilon_{\text{start}}}{\Delta\phi}. \tag{S 18}$$

From this we assign the new DTME phase value at the energy point $m$

$$\phi_{\text{out}}[m] = \phi_{\text{in}}[m] - \eta\nabla\mathcal{E} \tag{S 19}$$

where $\eta$ is the so-called learning rate. The same procedure is applied to the amplitude terms of the DTME vector and one iteration loop is completed when both amplitude and phase have been updated at each energy point. For the phase points we combined the gradient descent algorithm with Adam optimizer which modifies the value of $\nabla\mathcal{E}$ which is used in equation (S 19) resulting in a faster converging process. The details of the Adam optimizer can be found in the reference [24].

The reason why we apply a refinement of the complex $\tilde{d}$ using the stochastic gradient descent algorithm comes from the fact that in the mathematical derivation of $\tilde{d}$ a Taylor expansion up to the first order (equation (7)) has been used. This is a valid approximation as long as $\tilde{d}$ remains linear in the range $(k + \min(A)) \leq k \leq$ $(k + \max(A))$. In this approximation the intensity $I_0$ of the streaking field which determines the amplitude of the vector potential $A$ plays a significant role.

In Figure S 2 (a) we show the retrieved GD just using the equation (S 16) from simulated spectrograms characterized by three different IR intensities: $1\times10^{10}$ W/cm$^2$ (red symbols), $5\times10^{10}$ W/cm$^2$ (blue symbols) and $1\times10^{11}$ W/cm$^2$ (green symbols). For all the three spectrograms the same XUV pulse has been considered characterized by a spectral intensity centred at 32.5 eV, a chirp of -0.025 fs2 and FWHM bandwidth of ~5.5 eV. The fast oscillations of the theory curve (black line from reference [3]) are perfectly retrieved in the lowest IR intensity case. For $I_0$ of $5\times10^{10}$ W/cm$^2$ and $1\times10^{11}$ W/cm$^2$ a smoothening in the retrieved GD compared to the input curve is visible even though the qualitative trend is reproduced. The observed smoothening effect as $I_0$ increases is related to the truncation of the Taylor expansion of the complex DTME at the first order as $\tilde{d}$ is no longer linear over the range $(k + \min(A)) \leq k \leq (k + \max(A))$ for all $k$.

When we refine the result from equation (S 16) using a stochastic gradient descent algorithm we can overcome this limitation on the intensity $I_0$ of the IR field. We tested the result using the additional step with the gradient descent algorithm for three different IR intensities: $1\times10^{11}$ W/cm$^2$, $1\times10^{12}$ W/cm$^2$ and $1\times10^{13}$ W/cm$^2$. The results are shown in Figure S 2 (b) and are represented by the red, blue and green symbols, respectively. We observe a good agreement with the theoretical curve (black line) for all the IR intensities used to simulate the spectrograms. Note that for this specific analysis, the theoretical input curve (black line) has been extended at energies below 27.8 eV and above 40.5 eV in order to be able to simulate the spectrograms avoiding unphysical abrupt jumps in the simulated spectrogram amplitude (see black dashed line in Figure S 2).

The first step (up to equation (S 16)) of the algorithm is computationally inexpensive compared to the optimization step with the stochastic gradient descent algorithm. For this reason, running few iterations (1000 iterations) with the ACDC algorithm before considering the gradient descent algorithm improves the convergence process, reaching a lower final error when compared to the gradient descent algorithm only.

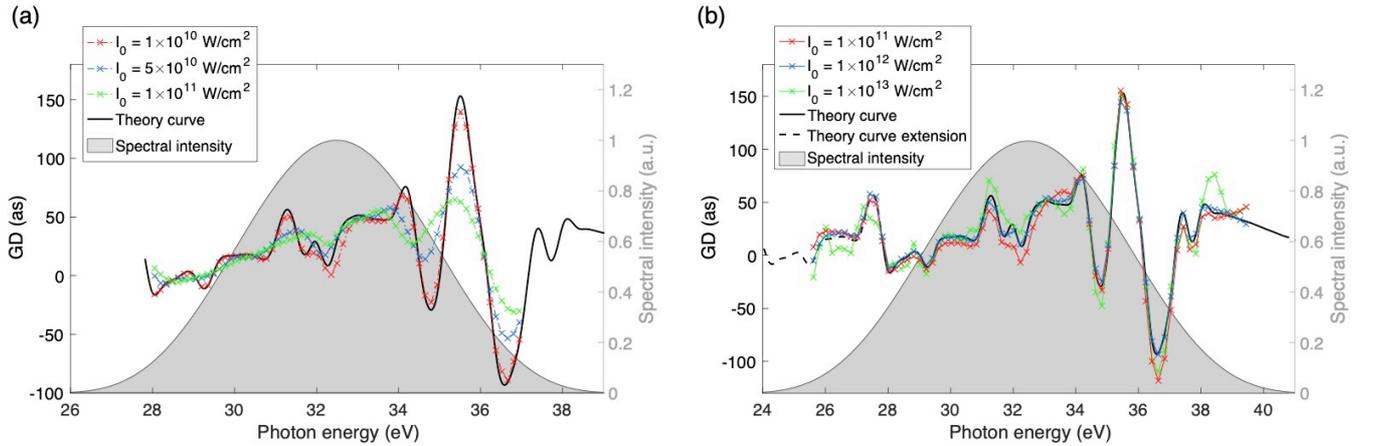

**Figure S 2.** (a) Retrieved GD using the ACDC algorithm up to the equation (S 16) for IR intensities of $1\times10^{10}$ W/cm$^2$ (red symbols), $5\times10^{10}$ W/cm$^2$ (blue symbols) and $1\times10^{11}$ W/cm$^2$ (green symbols). (b) Retrieved GD using the additional optimization step with stochastic gradient descent algorithm for IR intensities of $1\times10^{11}$ W/cm$^2$ (red symbols), $1\times10^{12}$ W/cm$^2$ (blue symbols) and $1\times10^{13}$ W/cm$^2$ (green symbols).

**Appendix C: Validation of the ACDC algorithm upon different streaking parameters**

In appendix A we demonstrated the ability of the ACDC algorithm to retrieve fast GD oscillations within few eV for different IR intensities. Here we want to test the reliability of the ACDC algorithm reconstructions for different XUV bandwidths and chirp values.

In Figure S 3 (a) we plot the results for XUV with FWHM bandwidths of ~3.8 eV (red line and red shaded area) and ~10.7 eV (blue line). We plotted for completeness also the green line representing the result for the FWHM bandwidth of ~7.4 eV already reported in Figure S 2 (b). We considered for this analysis a constant IR peak intensity of $1\times10^{12}$ W/cm$^2$ and XUV chirp of -0.025 fs$^2$. The features of the theory curve (black line) are perfectly resolved regardless of the XUV bandwidth considered.

In the same fashion, we tested the robustness of the retrieval algorithms to the XUV chirp. In Figure S 3 (b) we plot the retrieved GDs using values of XUV chirp: 0 fs$^2$ (red symbols), -0.015 fs$^2$ (blue symbols) and -0.025 fs$^2$ (green symbols). To simulate the spectrograms, we choose an XUV spectrum with FWHM bandwidth of ~7.4 eV (grey shaded area) and $I_0 = 1\times10^{12}$ W/cm$^2$. For chirped pulses and for the transform limited XUV pulse the retrieved GD reproduces accurately the theory curve (black line) making the two algorithms robust to different XUV chirp values. We found in general that phase changes near the central energy of the XUV pulse are more difficult to retrieve. For the transform-limited pulse, we observe a deviation of the retrieved GD from the theoretical curve of few points around the central energy of the XUV spectral intensity. Interestingly, we found that for the case of transform-limited pulses the solution is in fact not defined just at the central energy, while for the chirped pulses no significant error was observed.

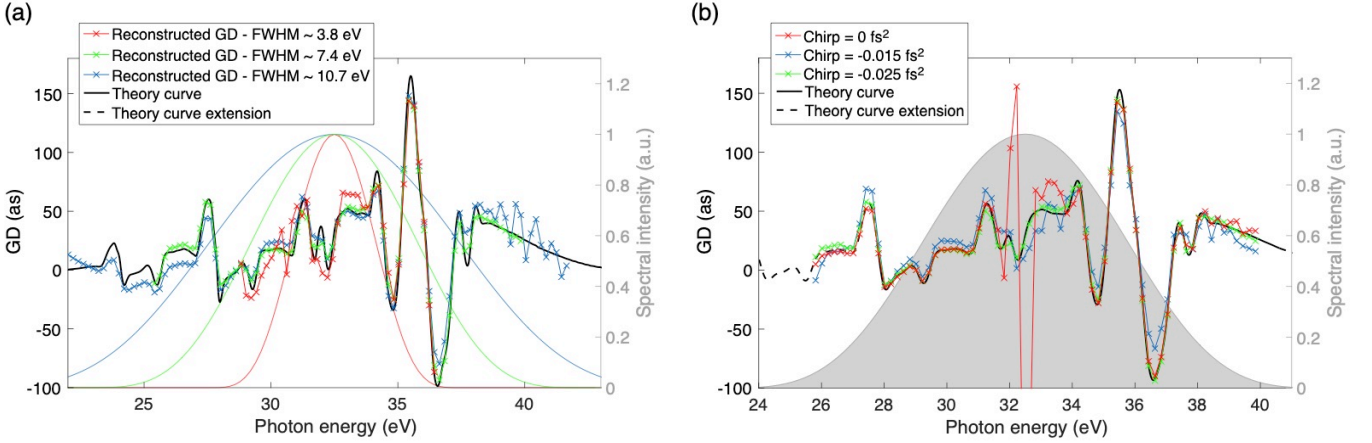

**Figure S 3.** (a) Red, green and blue symbols show the retrieved GD using the ACDC algorithm for XUV pulses centred at 32.5 eV, chirp of -0.025 fs$^2$ and FWHM bandwidths of ~3.8 eV (red line), ~7.4 eV (green line) and ~ 10.7 eV (blue line). (b) Retrieved GD using the ACDC algorithm for different XUV chirp values: 0 fs$^2$ (red symbols), -0.015 fs$^2$ (blue symbols) and -0.025 fs$^2$ (green symbols). The XUV spectral intensity is shown with the grey shaded area and is centred at 32.5 eV with FWHM bandwidth of ~ 7.4 eV. The IR field is kept constant for the three simulated spectrograms with an intensity of $1\times10^{12}$ W/cm$^2$.

### Appendix D: Noise analysis

Experimental measurements will always present a certain level of noise. Thus, testing the robustness of the combined algorithms against noise is important. It is worth highlighting the fact that, since the GD requires a differentiation of the retrieved dipole phase (see equation 2), so any noise-fluctuation will be amplified making the entire analysis challenging.

We define the signal-to-noise ratio (SNR) in the following way

$$\text{SNR} = \frac{\sqrt{\sum_{l,m} P[l,m]^2}}{\sqrt{\sum_{l,m} N[l,m]^2}} \qquad (S\ 20)$$

where $N[l,m]$ is the Poisson noise amplitude at the position of the $l$-time pixel and $m$-energy pixel.

Since we want to reconstruct experimental measurements, we considered the XUV and the IR field parameters similar to the experimental conditions. In particular, we considered both the XUV spectrum (see grey shaded area in Figure S 4) and IR streaking field ($I_0 = 3\times10^{12}$ W/cm$^2$) close to the experimental conditions.

In Figure S 4, the red line represents the reconstructed GD using the ACDC algorithm considering only a window of 2 fs in the pump-probe delay of the spectrogram for the noise free case. Such pump-probe delay window is enough to retrieve the GD information in the absence of noise. Adding artificial Poisson noise to the spectrogram, however, leads to a loss of information. An efficient way to recover such information is to reconstruct different pump-probe delay windows and average the output GD curves. In this way, the random noise contributions are averaged out while the GD information remains unaffected. To test this procedure, we added an amount of noise to the simulated spectrogram corresponding to a SNR level of ~7 according to equation (S 20). We reconstruct the obtained spectrogram after subdividing it in 19 different pump-probe delay windows, each having a width of 2 fs. Another approach would be to reduce the number of trace portions while broadening the pump-probe delay window. Assuming that the noise level is constant over the pump-probe delay scan, broadening the pump-probe delay window that we consider for the reconstruction will increase the normalized amount information with respect to the noise level that we provide to the algorithm. We confirmed this fact considering pump-probe delay of 6 fs of the noisy spectrogram and averaging 12 reconstructions. We could achieve the same qualitatively good result which is shown in Figure S 4. The SNR value of ~7 was chosen based on the estimated noise levels observed in our experimental data.

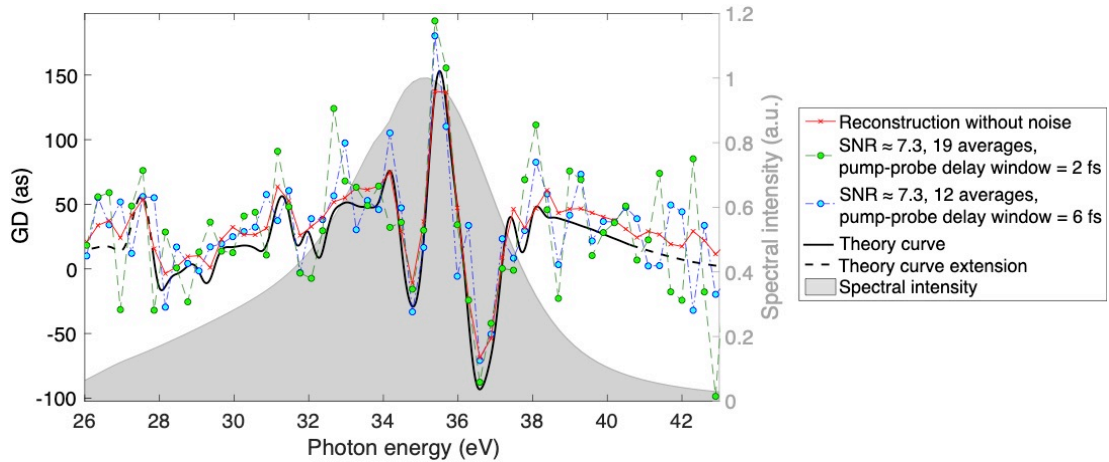

**Figure S 4.** Weighted average of 19 reconstructions on different pump probe delay windows of 2 fs of the noisy trace (green symbols) together with the weighted average of 12 reconstructions on different pump probe delay windows of 6 fs of the noisy trace (blue symbols). The theory curve used as input to generate the simulated spectrogram is shown by the black line.

In order to estimate the noise level that better approximate the experimental measurement we considered three noise levels which correspond to a SNR (defined by equation (S 20)) of approximately 23, 7.3 and 2.2. The original spectrogram together with those affected by the three noise levels are shown in Figure S 5.

In order to evaluate the SNR in the experimental measurement, we considered the photoemitted electron counts at each pump-probe delay bin in the Ne spectrogram. In Figure S 6(a) we plot with the blue line the photoemitted electron counts for the first delay bin. To isolate the signal, we applied a filtering procedure in the Fourier space where we eliminate the high noise components (see Figure S 6(b)). After filtering, the photoelectron count distribution results in the red line shown in Figure S 6(a). This photoelectron distribution gives an estimation of the signal level in our measurement which is then compared with the measured blue photoelectron distribution. Using this approach, we computed the SNR estimation of the experimental measurement performing the described filtering procedure for each pump-probe delay bins and averaging the SNR values obtained over all the delay steps.

This approach has been used to estimate the SNRs for the noisy simulated spectrograms shown in Figure S 5(b), (c) and (d). The resulting SNR values are 36, 12 and 4.4 respectively. In all the three cases the evaluated SNR with the described filtering procedure is overestimated if compared with the ones computed with equation (S 20) which leads to the values of 23, 7.3 and 2.2 respectively. However, the respective ratios are kept approximately the same.

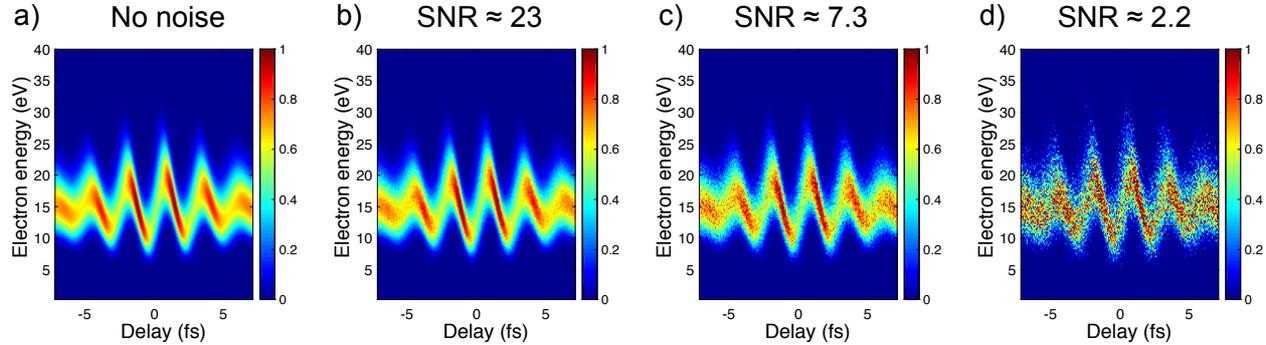

**Figure S 5.** (a) Simulated streaking spectrogram considered for testing the ACDC algorithm upon added Poisson noise. Three different noise levels have been considered and compared to the experimental case producing the spectrograms shown in (b) (SNR ≈ 23), (c) (SNR ≈ 7.3) and (d) (SNR ≈ 2.2).

Computing the SNR using the method described in this section on the experimental spectrogram we obtain the values of 13 and 11.7 for Ar and Ne, respectively. This brought us to assert that the SNR level in the experimental conditions analysed in this work can approximately be compared with SNR of about 7.3.

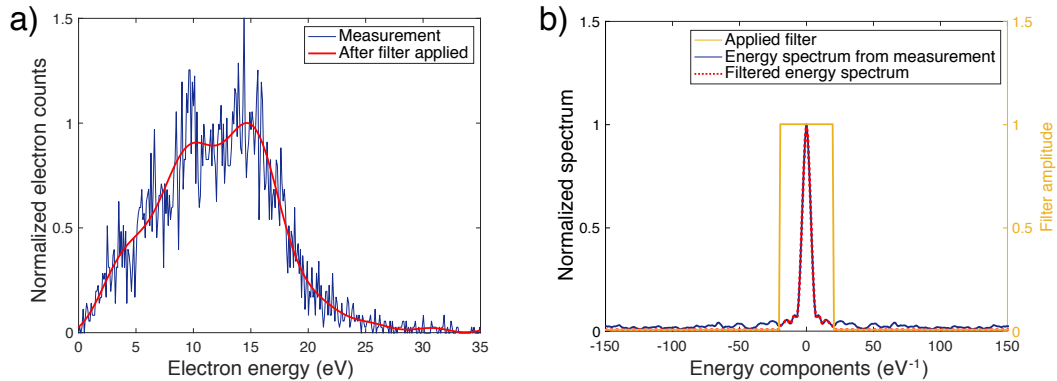

**Figure S 6.** Photoelectron counts distribution (blue line) normalized with respect to the filtered distribution (red line) corresponding to the first pump-probe delay step of the experimental Ne spectrogram. The energy components in the Fourier space of the measured photoelectron distribution are shown in (b) by the blue line. The red dotted line represents the portion of spectrum considered after the filter (yellow line) has been applied and which results in the red line in (a).

## Appendix E: Procedure for retrieving the dipole phase from experimental measurement using the ACDC algorithm

In section 5 we reported the dipole phase of Ar (Figure 4) reconstructed from the experimental streaking measurement shown in Figure 1 using the ACDC algorithm together with the computed GD difference between Ar and Ne (Figure 5). Here we describe in detail all the steps we followed to obtain the final result using the ACDC algorithm.

The first step consists in the characterization of both the XUV and IR fields. For this purpose, we considered the Ne spectrogram as a reference. We reconstructed, using VTGPA, three different pump-probe delay windows of 3 fs each (3-7 fs, 6-9 fs and 8.5-11.5 fs) of the Ne spectrogram, corresponding to the main oscillations in the streaking spectrogram. With the VTGPA we included the DTME of Ne in the reconstruction and extracted the information about the XUV and IR fields. We iterate for 3000 iterations and we considered the $\tilde{d}$ of Ne from reference [25].

To increase the accuracy of the streaking field reconstruction, an additional refinement only on the IR field has been performed after the 3000 iterations considering a wider pump probe delay window of 5 fs cantered around the original 3 fs pump probe delay window.

Since the Ar and Ne streaking spectrograms have been measured under the same experimental conditions thanks to the coincidence detection of the COLTRIMS apparatus we can assume that the XUV and IR fields characterized using the Ne spectrograms apply as well for Ar.

The characterized XUV and IR fields become then the inputs for the ACDC algorithm where now the target is the Ar spectrogram. Keeping fixed the XUV and IR fields characterized using VTGPA from the first pump-probe delay window (3-7 fs) of the Ne spectrogram, we selected 12 pump-probe delay windows of 2.5 fs (2.6-5.1 fs, 2.8-5.3 fs, 3.0-5.5 fs, 3.2-5.7 fs, 3.4-5.9 fs, 3.6-6.1 fs, 3.8-6.3 fs, 4-6.5 fs, 4.2-6.7 fs, 4.4-6.9 fs, 4.6-7.1 fs and 4.8-7.3 fs), around the 3-7 fs pump-probe delay window, for the Ar spectrogram which have been reconstructed by the ACDC algorithm. This procedure has been repeated similarly for the other two pump-probe delay windows used in the characterization step with the Ne spectrogram (6-9 fs and 8.5-11.5 fs) resulting in 36 total reconstructions using the ACDC algorithm. We considered 1000 iterations using the ACDC algorithm up to equation (S 16) as described in appendix A followed by 5000 iterations using the gradient descent algorithm described in appendix A. The final 36 reconstructions have been weighted averaged and resulted in the final dipole phase of Ar with its corresponding standard deviation shown in Figure 4.

**Appendix F: Acronyms**

| | |
|---|---|
| ACDC | absolute complex dipole transition matrix element reconstruction |
| APT | attosecond pulse train |
| COLTRIMS | Cold target recoil ion momentum spectrometer [18] |
| DTME | dipole transition matrix element |
| FROG-CRAB | frequency resolved optical gating for complete retrieval of attosecond burst [21], [22] |
| GD | group delay |
| IR | infrared |
| RABBITT | reconstruction of attosecond beating by interference of two-photon transition [8], [9] |
| SAP | single attosecond pulse |
| SFA | strong-field approximation |
| VTGPA | Volkov transform generalized projections algorithm [23] |
| XUV | extreme ultraviolet |

**Table 1.** List of the main acronyms used throughout this paper.